# acmqueue
## HOW WILL ASTRONOMY ARCHIVES SURVIVE THE DATA TSUNAMI?

**Astronomers are collecting more data than ever. What practices can keep them ahead of the flood?**

G. Bruce Berriman, NASA Exoplanet Science Institute, Infrared Processing and Analysis Center
Steven L. Groom, Infrared Processing and Analysis Center, California Institute of Technology

Astronomy is already awash with data: currently 1 PB (petabyte) of public data is electronically accessible, and this volume is growing at 0.5 PB per year. The availability of this data has already transformed research in astronomy, and the STScI (Space Telescope Science Institute) now reports that more papers are published with archived data sets than with newly acquired data.[17]

This growth in data size and anticipated usage will accelerate in the coming few years as new projects such as the LSST (Large Synoptic Survey Telescope), ALMA (Atacama Large Millimeter Array), and SKA (Square Kilometer Array) move into operation. These new projects will use much larger arrays of telescopes and detectors or much higher data acquisition rates than are now used. Projections indicate that by 2020, more than 60 PB of archived data will be accessible to astronomers.[9]

### THE TSUNAMI HAS ALREADY MADE LANDFALL

The data tsunami is already affecting the performance of astronomy archives and data centers. One example is the NASA/IPAC (Infrared Processing and Analysis Center) IRSA (Infrared Science Archive), which archives and serves data sets from NASA's infrared missions. It is going through a period of

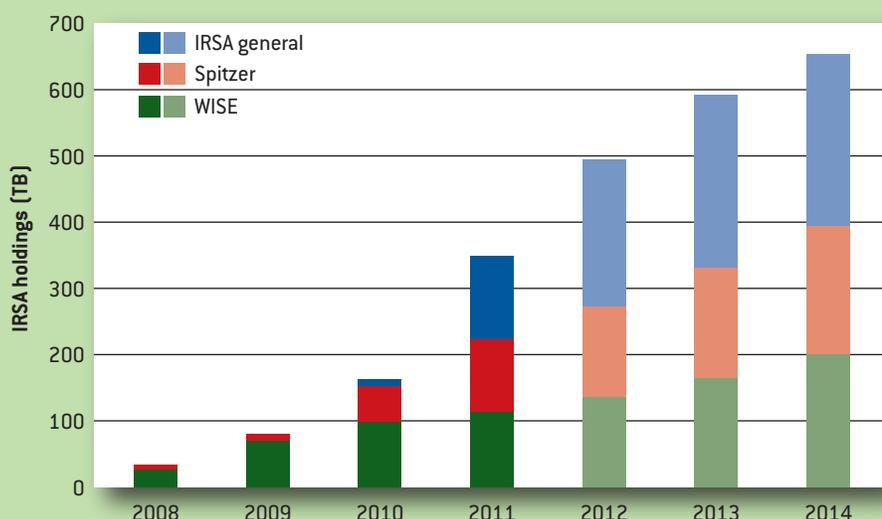

**FIGURE 1. Growth in the Scientific Data Holdings of IRSA, Projected to 2014**

*Chart courtesy of IRSA*

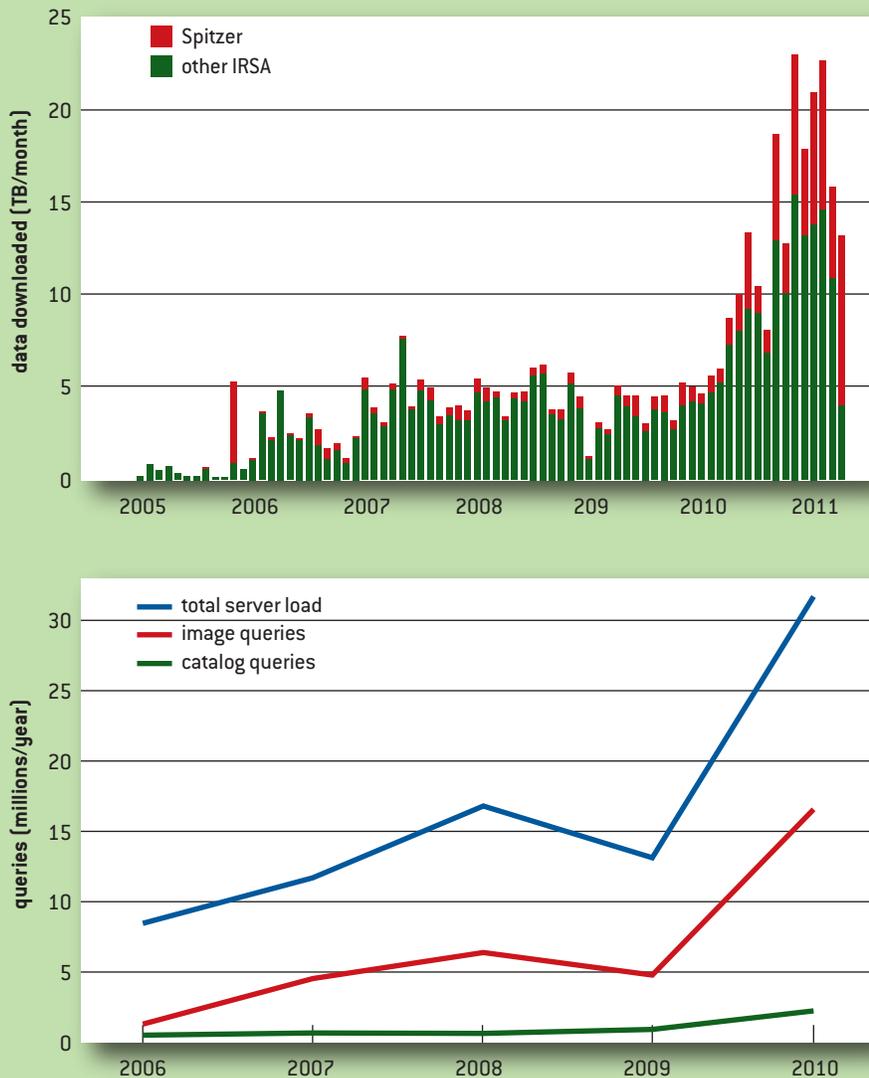

**FIGURE 2** Growth in Usage of IRSA from 2005 until the Beginning of 2011

*Chart courtesy of IRSA*

exceptional growth in its science holdings, as shown in figure 1, because it is assuming responsibility for the curation of data sets released by the Spitzer Space Telescope and WISE (Wide-field Infrared Survey Explorer) mission.

The volume of these two data sets alone exceeds the total volume of the 35-plus missions and projects already archived. The availability of the data, together with rapid growth in program-based queries, has driven up usage of the archive, as shown by the annual growth in downloaded data volume and queries in figure 2. Usage is expected to accelerate as new data sets are released through the archive, yet the response times to queries have already suffered, primarily because of a growth in requests for large volumes of data.

The degradation in performance cannot be corrected simply by adding infrastructure as usage increases, as is common in commercial enterprises, because astronomy archives generally operate on limited budgets that are fixed for several years. Without intervention, the current data-access and computing model used in astronomy, in which data downloaded from archives is analyzed on local machines, will break down rapidly. The very scale of data sets such as those just described will transform the design and operation of archives as places that not only make data accessible to users, but also support *in situ* processing of these data with the end users' software: network bandwidth limitations prevent transfer of data on this scale, and users' desktops in any case generally lack the power to process PB-scale data.

Moreover, data discovery, access, and processing are likely to be distributed across several archives, given that the maximum science return will involve federation of data from several archives, usually over a broad wavelength range, and in some cases will involve confrontation with large and complex simulations. Managing the impact of PB-scale data sets on archives and the community was recognized as an important infrastructure issue in the report of the 2010 Decadal Survey of Astronomy and Astrophysics,[5] commissioned by the National Academy of Sciences to recommend national priorities in astronomy for the coming decade.

Figure 3 illustrates the impact of the growth of archive holdings. As holdings grow, so does the demand for data, for more sophisticated types of queries, and for new areas of support, such as

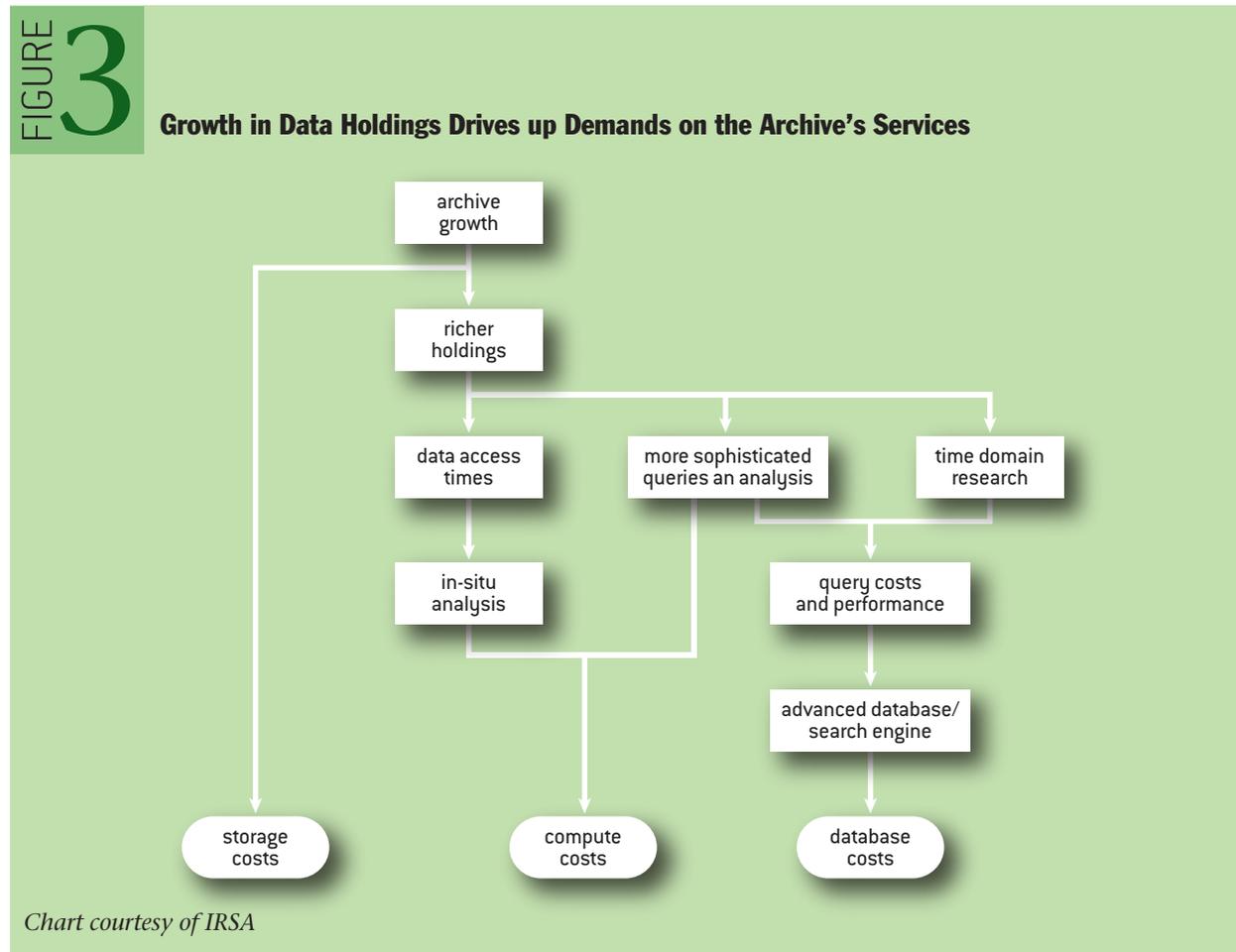

**FIGURE 3** **Growth in Data Holdings Drives up Demands on the Archive's Services**

*Chart courtesy of IRSA*

analysis of massive new data sets to understand how astronomical objects vary with time, described in the 2010 Decadal Survey as the "last frontier in astronomy." Thus, growth in holdings drives up storage costs, as well as compute and database costs, and the archive must bear all of these costs. Given that archives are likely to operate on shoestring budgets for the foreseeable future, the rest of this article looks at strategies and techniques for managing the data tsunami.

HOW TO KEEP THE TSUNAMI FROM ENGULFING US
At the Innovations in Data-intensive Astronomy workshop earlier this year (Green Bank, West Virginia, May 2011[14]) participants recognized that the problems of managing and serving massive data sets will require a community effort and partnerships with national cyber-infrastructure programs. The solutions will require rigorous investigation of emerging technologies and innovative approaches to discovering and serving, especially as archives are likely to continue to operate on limited budgets. How can archives develop new and efficient ways of discovering data? When should, for example, an archive adopt technologies such as GPUs (graphical processing units) or cloud computing? What kinds of technologies are needed to manage distribution of data time, computation-intensive data-access jobs, and end-user processing jobs?

This article emphasizes those issues that we believe need to be addressed by archives to support their end users in the coming decade, as well as those issues that affect end users in their interactions with archives.

INNOVATIONS IN SERVING AND DISCOVERING DATA
The discipline of astronomy needs new data-discovery techniques that respond to the anticipated growth in the size of data sets and that support efficient discovery of large data sets across distributed archives. These techniques must aim to offer data discovery and access across PB-sized data sets (e.g., discovering images over many wavelengths over a large swath of the sky such as the Galactic Plane) while preventing excessive loads on servers.

The VAO (Virtual Astronomical Observatory),[18] part of a worldwide effort to offer seamless international astronomical data-discovery services, is exploring such techniques. It is developing an R-tree-based indexing scheme that supports fast, scalable access to massive databases of astronomical sources and imaging data sets.[8] (R-trees are tree data structures used for indexing multidimensional information. They are commonly used to index database records and thereby speed up access times.)

In the current implementation, the indices are stored outside the database, in memory-mapped files that reside on a dedicated Linux cluster. It offers speed-ups of up to 1,000 times over database table scans and has been implemented on databases containing 2 billion records and TB-scale image sets. It is already in operation in the Spitzer Space Telescope Heritage Archive and the VAO Image and Catalog Discovery Service. Expanding techniques such as this to PB-scale data is an important next step.

Such custom solutions may prove more useful than adapting an expensive GIS (Geographical Information System) to astronomy. These systems are necessarily more complex than are needed in astronomy, where the celestial sphere is by definition a perfect sphere and the footprints on the sky of instruments and data sets are generally simple geometric shapes.

INVESTIGATIONS OF EMERGING TECHNOLOGIES
A growing number of investigators are taking part in a concerted and rigorous effort to understand how archives and data centers can take advantage of new technologies to reduce computational and financial costs.

Benjamin Barsdell et al.[1] and Christopher Fluke et al.[6] have investigated the applicability of GPUs to astronomy. Developed to accelerate the output of an image on a display device, GPUs consist of many floating-point processors. These authors point out that speed-ups of more than 100 times promised by manufacturers strictly apply to graphics-like applications; GPUs support single-precision calculations rather than the double-precision calculations often needed in astronomy; and their performance is often limited by data transfer to and from the GPUs. The two studies cited here indicate that applications that submit to "brute-force parallelization" will give the best performance with minimum development effort; they show that code profiling will likely help optimization, and provide a first list of the types of astronomical applications that may benefit from running on GPUs. These applications include fixed-resolution mesh simulations, as well as machine-learning and volume-rendering packages.

Others are investigating how to exploit cloud computing for astronomy. Applications that are best suited for commercial clouds are those that are processing- and memory-intensive, which take advantage of the relatively low cost of processing *under current fee structures*.[2] Applications that are I/O-intensive, which in astronomy often involve processing large quantities of image data, are, however, uneconomical to run because of the high cost of data transfer and storage. They require high-throughput networks and parallel file systems to achieve best performance.

Under current fee structures, renting mass storage space on the Amazon cloud is more expensive than purchasing it. Neither option offers a solution to the fundamental business problem that storage costs scale with volume, while funding does not. Any use of commercial clouds should be made after a thorough cost-benefit study. It may be that commercial clouds are best suited for short-term tasks, such as regression testing of applications and handling excessive server load, or to one-time bulk-processing tasks, as well as supporting end-user processing.

Implementing and managing new technologies always have a business cost, of course. Shane Canon[3] and others have provided a realistic assessment of the business impact of cloud computing. Studies such as these are needed for all emerging technologies.

Despite the high costs often associated with clouds, the virtualization technologies used in commercial clouds may prove valuable when used within a data center. Indeed, the CADC (Canadian Astronomy Data Center) is moving its entire operation to an academic cloud called CANFAR (Canadian Advanced Network for Astronomical Research), "an operational system for the delivery, processing, storage, analysis, and distribution of very large astronomical datasets. The goal of CANFAR is to support large Canadian astronomy projects."[10] To our knowledge, this is the first astronomy archive that has migrated to cloud technologies.[7] It can be considered a first model of the archive of the future, and consequently the community should monitor its performance.

The SKA has rejected the use of commercial cloud platforms. Instead, after a successful prototyping experiment, it proposes a design based on the open source Nereus V Cloud[19] computing technology, selected because of its Java codebase and security features. The prototype testbed used 200 clients at the University of Western Australia, Curtin University, and iVEC, with two servers deployed through management at a NereusCloud domain. The clients include Mac Minis and Linux-based desktop machines. When complete, "theskynet," as it has been called, would provide open access to the SKA data sets for professionals and citizen scientists alike.[11] The design offers a cheaper and much greener alternative to earlier designs based exclusively on a centrally based GPU cluster.

### COMPUTE INFRASTRUCTURE

Astronomy needs to engage and partner with national cyber infrastructure initiatives. Much of the infrastructure to optimize task scheduling and workflow performance and to support distributed processing of data is driven by the needs of science applications. Indeed, the IT community has adopted the Montage image mosaic engine[20] to develop infrastructure (e.g., task schedulers in distributed environments and workflow optimization techniques). These efforts have not, however, been formally organized, and future efforts may well benefit from such.

### CULTURAL CHANGES

There is at present no effective means of disseminating the latest IT knowledge to the astronomical community. Information is scattered across numerous journals and conference proceedings. To rectify this, we propose an interactive online journal dedicated to information technology in astronomy or even physical sciences as a whole.

Even more important is the need to change the reward system in astronomy to offer recognition for computational work. This would help retain quality people in the field.

Finally, astronomers must engage the computer science community to develop science-driven infrastructure. The SciDB database,[15] a PB-scale next-generation database optimized for science applications, is an excellent example of such collaboration.

### EDUCATIONAL CHANGES

An archive model that includes processing of data on servers local to the data will have profound implications for end users, who generally lack the skills not only to manage and maintain software, but also to develop software that is environment-agnostic and scalable to large data sets. Zeeya Merali[13] and Igor Chilingarian and Ivan Zolotukhin[4] have made compelling cases that self-teaching of software development is the root cause of this phenomenon. Chilingarian and Zolotukhin in particular present some telling examples of clumsy and inefficient design in astronomy.

One solution would be to make software engineering a mandatory part of graduate education, with a demonstration of competency as part of the formal requirements for graduation. Just as classes in instrumentation prepare students for a career in which they design experiments to obtain new data, so instruction in computer science prepares them for massive data-mining and processing tasks. Software has become, in effect, a scientific instrument.

The software engineering curriculum should include the principles of software requirements, design, and maintenance (version control, documentation, basics of design for adequate testing); how a computer works and what limits its performance; at least one low-level language and one scripting language, development of portable code, parallel-processing techniques, principles of databases, and how to use high-performance platforms such as clouds, clusters, and grids. Teaching high-performance computing techniques is particularly important, as the load on servers needs to be kept under control. Such a curriculum would position astronomers to develop their own scalable code and to work with computer scientists in supporting next-generation applications.

Curriculum designers can take advantage of existing teaching methods. Software Carpentry[16]

is an open source project that provides online classes in the basics of software engineering and encourages contributions from its user community. Frank Loffler et al.[12] described a graduate class in high-performance computing at Louisiana State University in which they used the TeraGrid to instruct students in high-performance computing techniques that they could then use in day-to-day research. Students were given hands-on experience at running simulation codes on the TeraGrid, including codes to model black holes, predict the effects of hurricanes, and optimize oil and gas production from underground reservoirs.

CONCLUSION

The field of astronomy is starting to generate more data than can be managed, served and processed by current techniques. This paper has outlined practices for developing next-generation tools and techniques for surviving this data tsunami, including rigorous evaluation of new technologies, partnerships between astronomers and computer scientists, and training of scientists in high-end software engineering engineering skills.

**LOVE IT, HATE IT? LET US KNOW**

feedback@queue.acm.org

**G. BRUCE BERRIMAN** is a senior scientist at IPAC (Infrared Processing and Analysis Center). He is the program manager for the Virtual Astronomical Observatory and project manager for the W. M. Keck Observatory Archive and was formerly the manager of the NASA/IPAC Infrared Science Archive. He obtained a Ph.D. in astronomy from Caltech and has performed research in cataclysmic variable stars, polarimetry, and brown dwarfs. He supported the Cosmic Background Explorer mission at Goddard Space Flight Center. He is active in investigating the applicability of emerging technologies to managing, serving, and processing massive astronomy data sets.
**STEVEN L. GROOM** is a systems engineer at IPAC and is manager of the NASA/IPAC Infrared Science Archive. He has an M.S. in computer science from the University of California Riverside and has worked with mass storage, parallel processing, and data archiving in the space sciences, as well as commercial applications.